\documentclass[useAMS, fleqn,usenatbib]{mnras}

\usepackage{newtxtext,newtxmath}
\usepackage[utf8]{inputenc}
\usepackage{amsmath}
\usepackage{amsfonts}
\usepackage{amssymb}
\usepackage{natbib}
\usepackage{graphicx}
\usepackage{aas_macros}
\usepackage[usenames,dvipsnames]{color}
\usepackage{xspace}
\usepackage{bm}
\usepackage{svn-multi}
\svnidlong{$LastChangedBy: n7216576 $}{$LastChangedRevision: 6212 $}{$LastChangedDate: 2019-05-08 02:54:49 +0100 (Wed, 08 May 2019) $}{$HeadURL: https://isssvn.ncl.ac.uk/svn/ngrs/ism/papers/CR_subgrid/paper_A.tex $}

\usepackage[normalem]{ulem} %

\usepackage{cancel} 

\newcommand\od{\mathrm{d}}
\newcommand\e{\mathrm{e}}
\newcommand\const{\mathrm{const}}
\newcommand\deriv[2]{\frac{\upartial#1}{\upartial#2}}
\newcommand\oderiv[2]{\frac{\od#1}{\od#2}}

\renewcommand{\vec}{\bm}
\newcommand{\red}[1]{#1}

\definecolor{webgreen}{rgb}{0,.5,0}
\definecolor{b}{rgb}{0,.2,0.45}
\definecolor{webbrown}{rgb}{.6,0,0}
\definecolor{purple}{rgb}{0.5,0,.5}

\DeclareMathOperator*{\dv}{d \!}
\renewcommand{\pi}{\upi}
\renewcommand{\partial}{\upartial}

\newcommand{\cm}{\,{\rm cm}}      
      
\newcommand{\pc}{\,{\rm pc}}

\newcommand{\s}{\,{\rm s}}      
  
\newcommand{\yr}{\,{\rm yr}}  

\newcommand{\GeV}{\,{\rm GeV}}  
\newcommand{\muG}{\,\mu{\rm G}}

\newcommand{\ecr}{\varepsilon}
\newcommand{\kperp}{\kappa_\perp}
\newcommand{\kpara}{\kappa_\parallel}
\newcommand{\keff}{\kappa_\text{eff}}

\newcommand{\kmin}{k_\text{min}}
\newcommand{\kmaxp}{k_\text{max}^\text{particle}}
\newcommand{\kmaxf}{k_\text{max}^\text{fluid}}
\newcommand{\kpeak}{\kappa_\text{peak}}
\newcommand{\tpeak}{t_\text{peak}}
\newcommand{\Fcrv}{{F}}
\newcommand{\CR}{{_\mathrm{cr}}}
\newcommand{\Brms}{B_{\mathrm{rms}}}
\newcommand{\RL}{R_{\mathrm{L}}}

\hypersetup{
 pdfauthor={A.~P.~Snodin, L.~F.~S.~Rodrigues},%
pdftitle={Fickian and non-Fickian diffusion of cosmic rays},
pdfkeywords={dif
fusion -- magnetic fields -- turbulence},
bookmarksnumbered=true,
}

\graphicspath{fig/}

\title[Fickian and non-Fickian diffusion of cosmic rays]
{Fickian and non-Fickian diffusion of cosmic rays}

\author[L.~F.~S.~Rodrigues, A.~P.~Snodin, G.~R.~Sarson, A.~Shukurov]{
Luiz~F.~S.~Rodrigues,$^1$ Andrew~P.~Snodin,$^2$
Graeme~R.~Sarson,$^1$ Anvar~Shukurov$^1$\thanks{
E-mail:
luiz.rodrigues@newcastle.ac.uk;
andrew.snodin@gmail.com;
g.r.sarson@newcastle.ac.uk;
anvar.shukurov@newcastle.ac.uk
}
\\
$^1$ School of Mathematics, Statistics and Physics, University of Newcastle,
Newcastle upon Tyne, NE1 7RU, UK \\
$^2$Department of Mathematics, Faculty of Applied Science, King Mongkut's University
of Technology North Bangkok, Bangkok 10800, Thailand
}
\date{Accepted for publication in MNRAS}
\pubyear{2018}
\pagerange{\pageref{firstpage}--\pageref{lastpage}}

\volume{{\rm in press}}
\begin{document}

\maketitle

\begin{abstract}
Fluid approximations to cosmic ray (CR) transport are often preferred to
kinetic descriptions in studies of the dynamics of the interstellar 
medium (ISM) of galaxies, because they allow simpler analytical 
and numerical treatments.
Magnetohydrodynamic (MHD) simulations of the ISM usually 
incorporate CR dynamics as an advection-diffusion equation for CR energy 
density, with anisotropic, magnetic field-aligned diffusion with the diffusive
flux assumed to obey Fick's law. We compare test-particle and fluid
simulations of CRs in a random magnetic field.
We demonstrate that a non-Fickian prescription of CR diffusion,
which corresponds to the telegraph equation for the CR energy density, can be 
easily calibrated to match the test particle simulations 
with great accuracy.
In particular, we consider a random magnetic field in the fluid simulation 
that has a lower spatial resolution than that used in the particle simulation
to demonstrate that an appropriate choice of the diffusion tensor can account 
effectively for the unresolved (subgrid) scales of the magnetic field.
We show that the characteristic time which appears in the telegraph equation can be
physically interpreted as the time required for the particles to reach a diffusive
regime and we stress that the Fickian description of the CR fluid is unable to
describe complex boundary or initial conditions for the CR energy flux.
\end{abstract} 


\label{firstpage}
\section{Introduction}

Cosmic rays (CRs) are an important ingredient of the interstellar medium
(ISM)  of galaxies, 
{with a typical energy density comparable to the magnetic energy \red{density} and the
turbulent gas kinetic energy \red{density}, and they contribute significantly to the support
of the galactic gaseous discs against gravity and drive galactic outflows 
\citep{Kulsrud2005}}.
Significant efforts have been made to include CRs in MHD simulations of the ISM
and assess their importance in the outflows 
\citep[e.g.,][]{Simpson2016,Girichidis2016,Holguin2018,Farber2018}
and galactic dynamos \citep[e.g.,][]{HanaszEA2009}.
The transport of CRs is usually quite simplified in these works, where the CR population
is modelled as a single fluid under the advection-diffusion approximation,
{with}
the diffusive
flux taken to follow Fick's law. 
CR diffusion is assumed  to be either isotropic,
anisotropic but with negligible diffusion
{perpendicular to the magnetic field},
or to have both parallel and perpendicular diffusivities 
\citep{SMP07,Zw13,GBS15,Zweibel2017}.
In all of these cases, constant values are taken for the diffusivities, with a parallel
{(or isotropic)}
component typically of order $\kpara=10^{28}\cm^2\s^{-1}$ and a perpendicular
diffusivity, $\kperp$, $10\text{--}1000$ times smaller. 

Meanwhile, a coherent theory of CR propagation and confinement has been
sought, mostly using quasi-linear theory 
 \citep[see][and references therein]{SMP07}
verified and extended using simulations of test particles
\citep[e.g.,][and references therein]{Snodin2016}.

One of {the} dynamical roles of CRs in the ISM is to provide an additional force due
to their pressure gradient that acts on the background gas. The CR particles
responsible for this effect gyrate about magnetic field lines with
a Larmor radius 
that is much smaller than scales resolved in modern ISM simulations. 
The CR fluid description should then account
for the effective CR dynamics at the scale resolved in the simulations. Here 
we consider CR diffusion in random magnetic fields which can be examined
via test-particle simulations where the relevant small scales are reasonably
well resolved. 
We devise simple numerical experiments to examine whether
anisotropic diffusion prescriptions with constant $\kpara/\kperp$, applied to
fluid simulations at a relatively low spatial resolution can capture the 
properties of test particle propagation in simulations at higher resolution.
In particular, we compare the standard anisotropic Fickian diffusion of CRs 
with a non-Fickian one that leads, instead of the diffusion equation, to the 
telegraph equation for the number (or energy) density of the particles.
The telegraph equation contains an additional time scale $\tau$ which accounts 
for the finite propagation speed of the CR distribution 
\citep[e.g.,][]{Bakunin} and allows for the differentiation between the early 
ballistic behaviour of the particles and the subsequent diffusive CR spread.
The telegraph equation is also a convenient way to avoid numerical problems close to
singular magnetic X-points, {and may also allow for larger time steps in an 
explicit numerical scheme than that with the Fickian approach} 
\mbox{\citep{Snodin2006}}.
\citet{LN16} compare numerical solutions of fluid equations for cosmic ray
propagation that result from various approximations to the Fokker--Planck 
equation, including the telegraph equation, and confirm the relevance of this
approximation. Our goal is to compare test-particle simulations of cosmic ray 
propagation with the telegraph-equation approximation to derive optimal 
parameters of the latter that can inform sub-grid models of cosmic ray 
propagation in \red{a} comprehensive MHD simulation of the multi-phase interstellar
medium.
We discuss the dependence of the CR diffusion parameters on the CR properties, on the
test particle magnetic field power spectrum and on the resolution of the fluid 
simulation.
We describe our numerical experiments with test particle and fluid descriptions 
in Sections~\ref{sec:test} and~\ref{sec:fluid}, respectively. The general 
results and a
proposed calibration procedure for the diffusion parameters are presented in
Sections~\ref{sec:results_test} and~\ref{sec:results_fluid}.
In Section \ref{sec:conclusion}, we discuss the
results further and summarise our conclusions.

\section{Numerical experiments}

\subsection{Test particle simulations}
\label{sec:test}

We use test particle simulations of CR propagation in a random magnetic field 
as a reference to which fluid simulations can be
compared and calibrated. These simulations are very similar to those performed by
\citet{Snodin2016}. For a direct comparison with fluid simulations, we 
construct a
random, time-independent, periodic, magnetic field $\vec{B}(\vec{x})$ on a 
uniform Cartesian
mesh of side length $L$ and resolution $1280$ points in each direction
\citep[the \emph{discrete model} described by][]{Snodin2016}.
We use an isotropic magnetic field with a power spectrum $M(k)\propto k^{s}$, 
over the range
$\kmin \le k \le \kmaxp$, with $\kmin = 2{\rm \pi}/L $ and
$\kmaxp = 640{\rm \pi}/L = 320 \kmin$, and consider the cases $s=-5/3$, and 
$s=-2$.
Within the simulation volume $L^3$, we placed $2.1\times10^7$ test particles of 
a given energy,
initially distributed in space so that their number density follows
\begin{equation}
  \label{eq:ncr}
	n(x,y,z) \propto \exp\left[-\frac{x^2}{2 \sigma_0^2}\right], \quad (x,y,z) \in [-L/2,L/2]^3\,,
\end{equation}
being non-uniform in the $x$-direction but homogeneous in the other directions.
The {half-width $\sigma_0$} was chosen to correspond to approximately
71 mesh points ($0.0552 L$).
The initial velocities of the particles were distributed isotropically.
For each particle, we then solved the equation of motion
\begin{equation} \label{eq:NewtonLorentz}
	\ddot{\vec{x}} = \alpha \dot{\vec{x}} \times \vec{B}[\vec{x}(t)]\,,
\end{equation}
where $\vec{x}(t)$ is the particle position, $\alpha=q L \Brms /(\gamma m c v_0)$,
with $q$ the particle charge, $m$ the particle mass, $\gamma$ the particle Lorentz
factor, $c$ the speed of light, $v_0$ a reference speed, and $\Brms$ the
\red{root-mean-square}
magnetic field strength. Here we neglect electric fields, so that the
particle energy is conserved.
We take $\alpha=320$, and $v/v_0=1$ for each particle, so that the 
dimensionless Larmor radius,
$R_{\rm L}/L = 1/320$, is somewhat smaller than the scales resolved in the fluid
simulations discussed below, but also a few times larger than the test-particle
grid resolution.
Typically, each particle trajectory was integrated up to a maximum dimensionless time
of $t=2.8$, corresponding to approximately $142 t_{\rm L}$,
where $t_{\rm L}=2\pi R_{\rm L}/v_0$ is the  Larmor time scale.

\subsection{Fluid description of CR propagation}
\label{sec:fluid}

We compare the test-particle results with simulations where
the CRs are approximated as a fluid governed by the equations
\citep{Snodin2006}
\begin{equation}
    \label{eq:ecr2}
    \frac{\partial\ecr}{\partial t}
        =
    -\nabla\cdot\vec{\Fcrv}\,
    \;,\qquad
    \tau \frac{\partial \Fcrv_i}{\partial t}=
    -\kappa_{ij}\nabla_j\ecr
    - \Fcrv_i\, ,
\end{equation}
where $\ecr$ is the CR energy density, $\vec\Fcrv$ is the associated energy 
flux and $\kappa_{ij}$ is a diffusion tensor given by
\begin{equation}\label{eq:ecr10}
  \kappa_{ij}=\kperp\delta_{ij}+(\kpara-\kperp)\hat{B}_i \hat{B}_j\,,
\end{equation}
where $\hat{\vec{B}}=\vec{B}/|\vec{B}|$ is the unit vector along the magnetic field.
Here the diffusivity along magnetic field lines is usually taken to be much larger than the
diffusivity perpendicular to field lines, $\kpara\gg \kperp$.
The case $\tau=0$ corresponds to anisotropic Fickian diffusion, which we refer
to as the \emph{Fickian} prescription. The case $\tau=0$ leads to the CR 
distribution spreading at an infinite speed,
so that an initially spatially localised CR energy distribution will be 
redistributed over all space available after any finite time. This is a 
well-known artefact of the diffusion approximation.
When $\tau\neq 0$, these equations correspond to a generalised form of the 
telegraph equation for $\ecr$,
which we shall refer to as the \emph{Telegraph} prescription.

\subsubsection{CR propagation and the telegraph equation}\label{CRTEq}
The telegraph equation for the
CR distribution function has been studied in several previous works 
\citep[e.g.,][]{GombosiEA1993,LitvinenkoSchlickeiser2013,TL16}, including its 
connection to a hyperdiffusive equation that arises more
naturally in higher-order expansions of the energy flux \citep{Malkov2015}. 
The main reason for using this prescription has
been to capture the non-diffusive behaviour at short times, corresponding to (superdiffusive)
ballistic particle motion, followed by a transition to the diffusive regime.
For the CR distribution function, the time scale $\tau$ can be connected with 
the
pitch angle diffusivity \citep{LitvinenkoSchlickeiser2013}.
One difference from the Fickian case is that the maximum signal propagation speed
(parallel to magnetic field lines) is reduced to $(\kpara/\tau)^{1/2}$.

It has been claimed that the telegraph equation is not suitable for
modelling CR propagation as it does not conserve the total number of particles 
or, equivalently, their total energy \citep{Malkov2015}; see also 
\citet{TL16}. However, this assertion seems to be based on misunderstanding. 
Indeed, consider the one-dimensional telegraph equation for the energy density 
of particles $\ecr(x,t)$ in $|x|<\infty$:
\begin{equation}\label{TEq}
\tau\deriv{^2\ecr}{t^2}+\deriv{\ecr}{t}+V\deriv{\ecr}{x}=\kappa\deriv{^2\ecr}{x^2}\,,
\end{equation}
where $V$ is the advection speed, assumed to be constant for simplicity, and
$\kappa$ is the diffusivity. Adopting natural boundary conditions
$\ecr=0$ and $\partial\ecr/\partial x=0$ at $x=\pm\infty$,
the total energy of
the particles $E(t)=\int_{-\infty}^\infty\ecr(x,t)\,\od x$ can easily be shown 
to satisfy
\begin{equation}\label{TEqNt}
\oderiv{E}{t}+\tau\oderiv{^2E}{t^2}=0\,,
\end{equation}
which solves to yield
\begin{equation}\label{TEqN}
E=C+K\e^{-t/\tau}\,,
\end{equation}
where $C$ and $K$ are constants. Since the telegraph equation is of second 
order in $t$, two initial conditions for $E(t)$ have to be specified:
$E(0)=E_0$ at $t=0$ for the initial energy and $\partial E/\partial t=0$ at 
$t=0$. Then the only admissible solution is
$E=E_0=\const$, expressing the conservation of the total energy (and 
similarly for the total number of particles). Other authors disregard the 
second initial condition, which seems to be the cause of confusion regarding 
the conservation of $E$, as they then analyse an incomplete solution 
$E=E_0+K(\e^{-t/\tau}-1)$.

\begin{figure*}
  \centering
  \includegraphics[width=0.98\columnwidth]{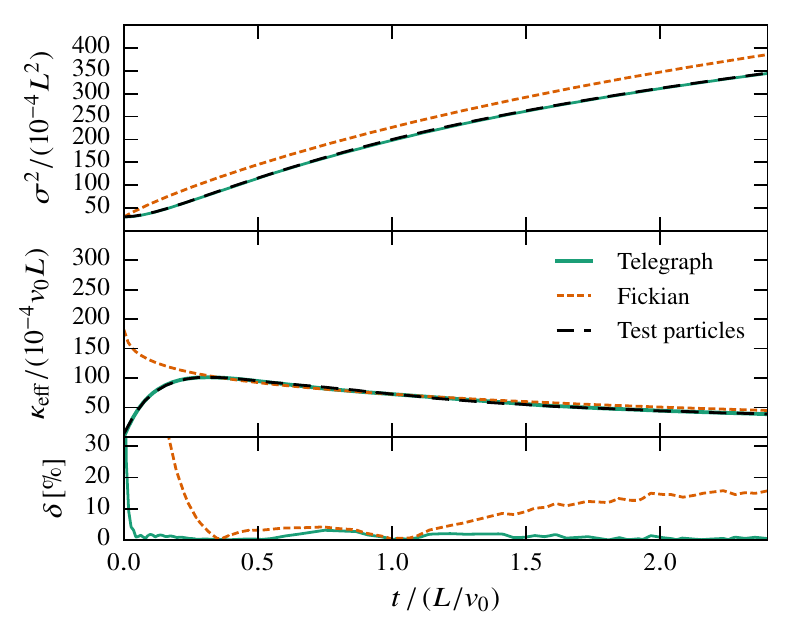}\quad\quad
  \includegraphics[width=0.98\columnwidth]{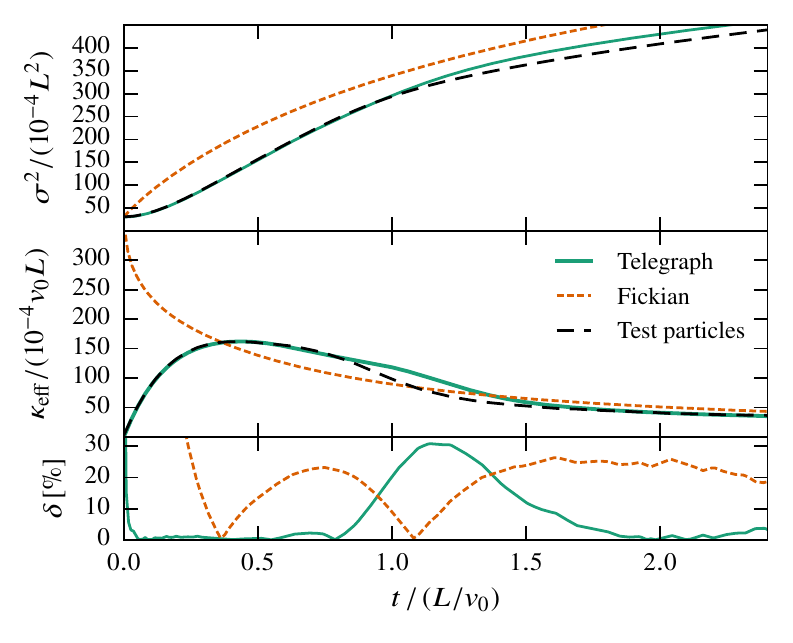}
  \caption{Comparison between the results of test particle simulations and
           fluid simulations using the Telegraph and Fickian prescriptions
           (with the same choice of calibrated diffusivities $\kpara$ and 
           $\kperp$)
           for a magnetic field with
           $s=-5/3$, in the left-hand side column,
           and $s=-2$ on the right.
           The top panels show time evolution of the variance $\sigma^2$ of the 
           CR distribution along the $x$-axis, the middle panels show the time 
           evolution of the effective
           diffusivity, $\keff=\tfrac12 \partial \sigma^2/\partial t$, and 
           bottom panels show the relative difference in $\keff$,
           $\delta=(\kappa_{\text{eff},\text{f}} 
           -\kappa_{\text{eff},\text{p}})/\kappa_{\text{eff},\text{p}}$,
with indices `f' and `p' referring to the fluid and test particle descriptions.
           \label{fig:variance}}
  \centering
\end{figure*}

\subsubsection{Fluid simulations}\label{FS}

We solve equations~\eqref{eq:ecr2} and \eqref{eq:ecr10} using the 
\textsc{Pencil Code}\footnote{\url{http://pencil-code.nordita.org/}}
\citep{PencilCode},
a code for non-ideal MHD using sixth-order spatial finite differences.
The simulation domain comprises a box of $80^3$ grid points, with periodic boundary
conditions.\footnote{We examined the effect of other choices of boundary 
conditions, and the effect proved
negligible for this particular setup, since the width of the CR distribution 
(initially set to $2\sigma_0$)
is much smaller than $L$
throughout the simulation. }
The magnetic field is the same as that used in the
test particle simulation, but with the smallest scales removed by truncating
the spectrum $M(k)$ at $\kmaxf=40\pi/L=\kmaxp/16$, resulting in a magnetic 
field specified on a coarser grid. 
As in the test-particle simulations, the magnetic field does not evolve
during the simulation.

The choice of a much coarser resolution for the magnetic field in the fluid
simulation reflects the fact that it is usually not feasible to simultaneously resolve
the scales of interest of a particular astrophysical problem and the Larmor radius of
the relevant CR population. For example, the typical resolution in simulations 
of milti-phase supernova-driven ISM is of
order of a few $\pc$, while the Larmor radius is of order 
$10^{-6}$--$10^{-5}\pc$ for cosmic ray energies of $1\text{--}10\GeV$ in a 
$5\muG$ magnetic field.
Thus, any effects arising from the
unresolved small scales, $k > \kmaxf$, need to be accounted for
through an adequate choice of the parameters that control CR
diffusion, i.e., the diffusivities $\kpara$, $\kperp$, and, in the Telegraph
case, the characteristic timescale $\tau$.

We chose an initial condition for the CR energy density consistent with the
particle distribution in Eq.~\eqref{eq:ncr} by assuming a single population of 
CRs, i.e.,
$\ecr(x,y,z) =n(x,y,z) E\CR$, where $n$ is the particle number density and 
$E\CR$ is the energy of an individual particle. In the fluid simulation, the 
initial half-width of the CR distribution, $\sigma_0$, corresponds to 
approximately {$4.4$ mesh points, consistent with the particle simulations}.

While in the Fickian prescription the initial condition for $\vec\Fcrv$ is fully
determined by the initial condition for $\ecr$, the Telegraph prescription 
requires a separate initial condition for $\vec\Fcrv$.
In agreement with the initial condition used for the particles, whose initial
velocities have random, isotropic directions, we impose the zero flux initial 
condition, $\vec\Fcrv=\vec{0}$.

\section{Results}

\subsection{Test particle simulations}
\label{sec:results_test}
Results of the test particle simulations are shown as dashed
curves in Fig.~\ref{fig:variance},
where the magnetic field has spectral index $s=-5/3$ (left-hand column) and $s=-2$
(right-hand column).
In the upper panels, we show the variance $\sigma^2$ of the CR number density
distribution along the $x$-axis as a function of time.
The \emph{global effective diffusivity} 
$\keff\equiv\frac{1}{2}\dv\sigma^2/\dv t$ is shown in the middle panels.
The variance $\sigma^2$ of the test  particle distribution
deviates strongly from the
simple linear dependence on time that would be expected in the case of an 
isotropic Fickian diffusion. This is also visible in the variation of $\keff$ 
with time, which first increases sharply, reaches a peak, and then decays to an 
asymptotic value.
This behaviour reflects the transition from ballistic to diffusive behaviour
and is characterised by the time interval $\tpeak$ after which $\keff$ reaches
a maximum $\kpeak$. The transition is sensitive to the spectral index of the 
random field,
with $\tpeak\approx 0.31L/v_0$ 
for $s=-5/3$ and
$\tpeak\approx0.43 L/v_0$ 
for $s=-2$.

\begin{figure*}
  \centering
  \includegraphics[width=0.99\columnwidth]{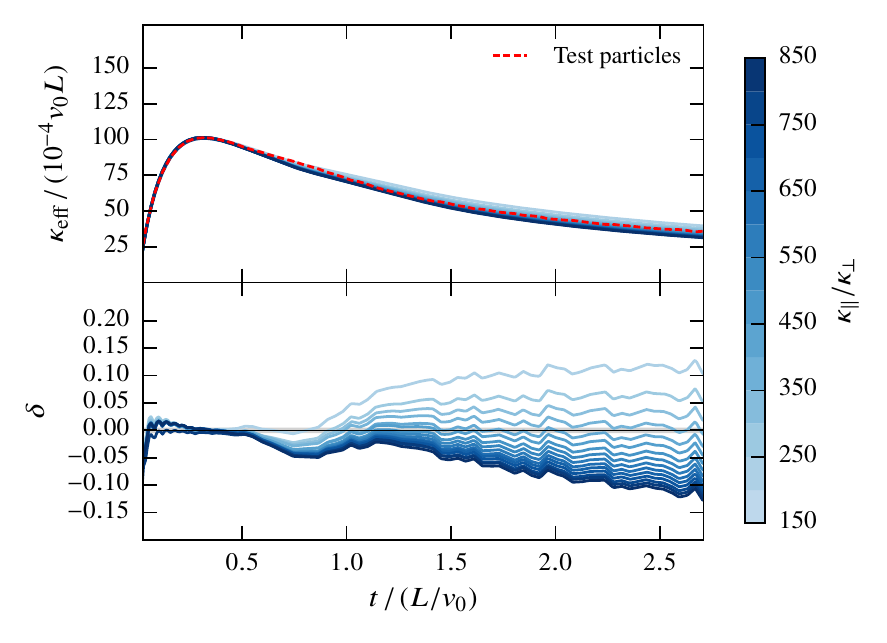}\quad\quad
  \includegraphics[width=0.99\columnwidth]{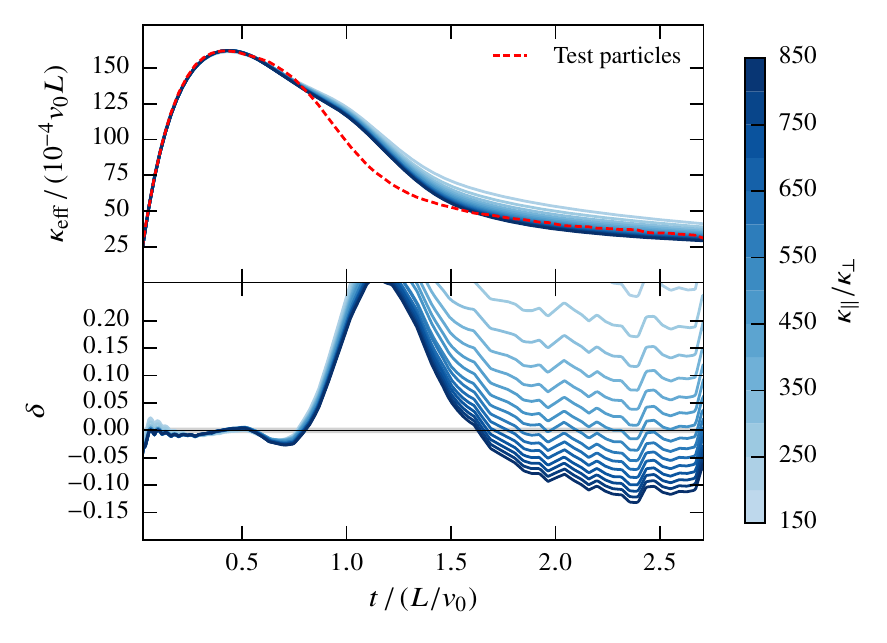}
  \caption{Impact of the choice of the ratio $r=\kpara/\kperp$ for a magnetic 
  field with
           $s=-5/3$ (left-hand column),
           and $s=-2$ (right-hand column).
           The top panels show the time evolution of the effective global 
           diffusivity $\keff$ for properly calibrated Telegraph models with 
           different choices of $r$ (colour-coded, with the colour bar shown 
           oon the right of each column).
           The bottom panels show the fractional error in $\kappa_\text{eff}$.
         \label{fig:ratio}}
\end{figure*}
\subsection{Fluid simulations}
\label{sec:results_fluid}
We found that it is possible to reproduce the behaviour of the test particle simulation
with the Telegraph fluid simulation. For a fixed value of $r=\kpara/\kperp$, we found
that $\tpeak$ and $\kpeak$ are controlled by $\tau$ and $\kpara$, respectively.
\red{Therefore,}
the fluid simulation can be calibrated using the following simple iterative
procedure, where the iteration steps are labelled with index $i$:
\begin{equation}
        \kpara^{(i+1)} = \kpara^{(i)} \left.
                        \frac{\kpeak^\text{particle}}{\kpeak^{(i)}}\right.
        \;,\quad
        \tau^{(i+1)} = \tau^{(i)}\left.\frac{\tpeak^\text{particle}}{\tpeak^{(i)}}\right.
        \label{eq:calibration}\,.
\end{equation}
\red{Thus, one runs a fluid simulation for given $\kpara^{(i)}$ and $\tau^{(i)}$, and from this run $\kpeak^{(i)}$ and $\tpeak^{(i)}$ are computed. Then, in
the next iteration, the fluid simulation is re-run using
$\kpara^{(i+1)}$ and $\tau^{(i+1)}$;
note that for this particular simulation setup the computational cost of each run is very low.
}
The iteration is performed
until both $\kpeak$ and $\tpeak$ differ from
$\kpeak^\text{particle}$ and $\tpeak^\text{particle}$, respectively,
by less than $0.5$ per cent.
The iterations typically converge to this accuracy in 2--4 steps.

\begin{figure}
  \includegraphics[width=\columnwidth]{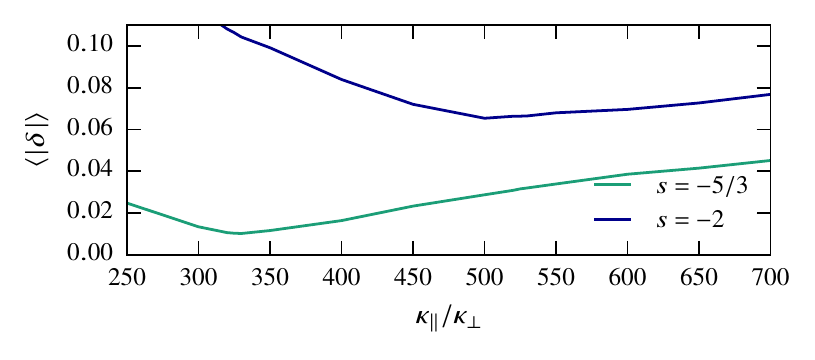}
  \caption{
           The time-averaged fractional error in $\keff$ as a function of the 
           ratio of CR diffusivities for two choices of the spectral index
of the magnetic spectrum as specified in the legend.
           }
  \label{fig:ratio_delta}
\end{figure}

In Fig.~\ref{fig:ratio}, each of the curves is obtained with a different choice 
of $r$,
calibrated using Eq.~\eqref{eq:calibration}. This ratio controls the
slope of $\keff(t)$ at late times, therefore controlling the asymptotic value of
$\keff$. The relative
difference between $\keff$ obtained from the fluid and particle simulations,
{$\delta=(\kappa_{\text{eff}}^{\text{fluid}} 
-\kappa_{\text{eff}}^{\text{particle}})/\kappa_{\text{eff}}^{\text{particle}}$},
is shown in the bottom panel.
We use the time averaged value $\langle\delta\rangle$ as a goodness-of-fit 
measure, which is shown in Fig.~\ref{fig:ratio_delta}. We find that the 
ratio $r=\kpara/\kperp$ giving the best fit to the test-particle results
depends on the spectrum of the
magnetic field, with $r\approx325$ for $s=-5/3$, and $r\approx525$ for $s=-2$.

\begin{table}
  \caption{The best-fit parameters used in th fluid simulations shown in
           Fig.~\ref{fig:variance}, values of $\tpeak$ and
           $\kpeak$, and their relation to the model 
           parameters.\label{tab:parameters}
           }
  \begin{center}
    \begin{tabular}{ccc}
    \hline
                                    & $s=-5/3$  & $s=-2$    \\
    \hline
        $\kpeak/(v_0L)$             & 0.010    & 0.016    \\
        $\kpara/(v_0L)$             & 0.054    & 0.109    \\
        $\kperp/(v_0L)$             & 0.00017   & 0.00021   \\
        $\kpara/\kperp$             & 325       & 525        \\[4pt]
        $\tpeak/(L/v_0)$            & 0.31      & 0.43      \\
        $\tau/(L/v_0)$              & 0.17      & 0.33      \\[4pt]
        $\kpara/\kpeak$             & 5.35   & 6.70         \\
        $\tau/\tpeak$               & 0.55   & 0.77         \\
        $(\kpara/\tau)^{1/2}/v_0$    & 0.56   & 0.57  \\
        \hline

    \end{tabular}
  \end{center}
\end{table}

The best fit parameters are summarised in Table~\ref{tab:parameters} and
correspond
to the solid green curves in Fig.~\ref{fig:variance}.
The dashed orange curves in those figures show the Fickian prescription for the same
choice of $\kpara$ and $\kperp$. The lower panels of those figures show $\delta$ for
both fluid prescriptions compared with the test particle simulation.
\red{For $s=-5/3$, the effective diffusivity, $\keff$ of the CR distribution resulting
from the calibrated Telegraph simulation
differs from that in the test particle simulation by at most
$2$ per cent at any time.
For $s=-2$ we find similarly small differences in $\keff$
for $t\lesssim 0.8L/v_0$ and $t\gtrsim1.7L/v_0$, but $\delta$ increases up to
$\sim30$ per cent for $t\approx1.1 L/v_0$. This leaves an imprint in the variance at
later times (top right panel of Fig.~\ref{fig:variance}).
Albeit small, such a feature occurs independent of the specific parameter choice
(see right panel of Fig.~\ref{fig:ratio}).
}
The Fickian solution does not capture the ballistic phase of particle 
propagation. However, at sufficiently
large times, the effective diffusivity in the Fickian case is in good
agreement with both the Telegraph and test particle results.
\red{In considering these differences, it is perhaps worth noting that the diffusion coefficient is typically unknown by almost a factor of two}.

\begin{figure*}
  \centering
  \includegraphics[width=0.98\textwidth]{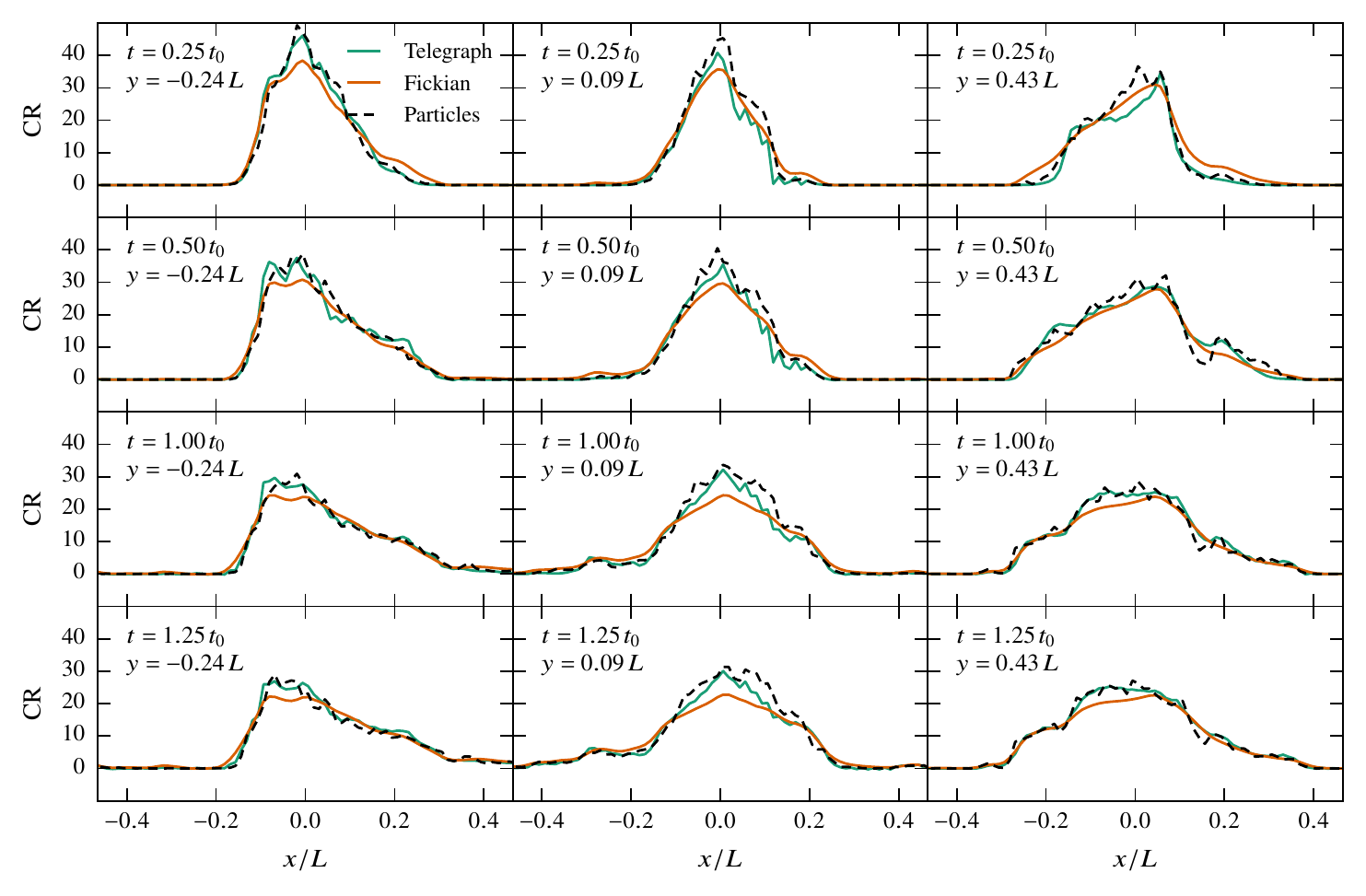}
  \caption{Time evolution (top to bottom) of the CR distribution along the 
  $x$-axis obtained by integrating the results over parallelepipeds extended 
  along the $x$-axis and $0.05 L$ in width along $y$ and $z$ passing through 
  arbitrarily chosen positions in $y$ and $z$ (left to right).
           The vertical axis shows, {in arbitrary units,}
	   the total number of particles and their energy in the integration 
	   domain. The higher accuracy of the Telegraph model is apparent.}
  \label{fig:slices}
\end{figure*}

The calibrated Telegraph model is able to reproduce local
features, as shown in Fig.~\ref{fig:slices}, where slices through the test
particle simulation box are compared with the corresponding slices of the
fluid simulations at various times and locations. Both the asymmetry in the
distribution and small-scale variations are reproduced better by the 
Telegraph model.

\subsection{Variability of $\tpeak$ in test particle simulations}\label{sec:testpeak}
\begin{figure}
  \centering
  \includegraphics[width=\columnwidth]{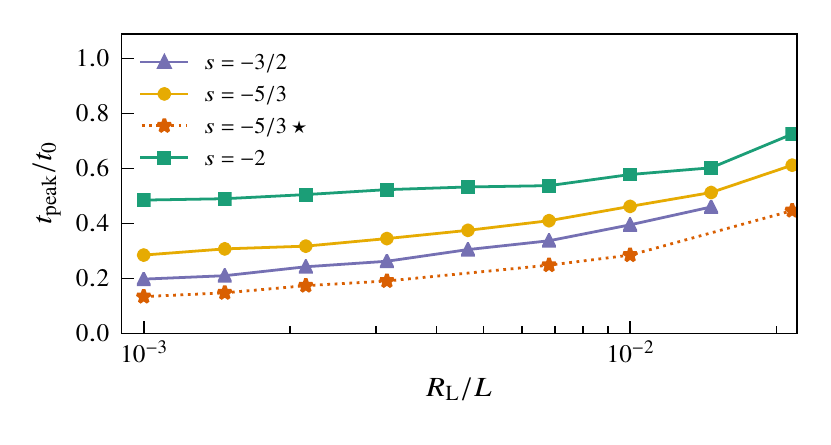}
	\caption{Dependence of $t_{\rm peak}$ on the Larmor radius $\RL$ (expressed
                 in units of $L$) in the test particle simulations,
                 for different choices of the magnetic spectral index $s$.
                 The data points shown with stars are obtained with a CR 
                 self-scattering model discussed in the text.
                 These results show that an adequate choice of the parameter 
                 $\tau$ needs to correctly account for the energy of the 
                 modelled CR population and the form of the (subgrid) magnetic 
                 spectrum.
                \label{fig:particleTpeak}
                (Data shown in Table~\ref{tab:particleTpeak}.)
                }
\end{figure}

\begin{table}
\newcommand{\mc}[3]{\multicolumn{#1}{#2}{#3}}
\begin{center}

\red{\caption{\label{tab:particleTpeak}
              Larmor radius and $\tpeak$ for different choices of magnetic spectral
              index, $s$. The star indicates the run with CR self-scattering model
              (see text).
              }}
 \begin{tabular}{c|cccc}
 \hline
 & \mc{4}{c}{$t_{\rm peak}/t_0$}\vspace{0.2em}\\
 $10^3 R_\text{L}/L$ \hspace{1em}                   &   $s=-3/2$ &   $s=-5/3$ &   $s=-5/3 \;\star$ &   $s=-2$ \\\hline
            1.00    &   0.20   &   0.28   &   0.13   &   0.48   \\
            1.47    &   0.21   &   0.31   &   0.15   &   0.49   \\
            2.15    &   0.24   &   0.32   &   0.17   &   0.51   \\
            3.16    &   0.26   &   0.34   &   0.19   &   0.52   \\
            4.64    &   0.30   &   0.38   &          &   0.53   \\
            6.81    &   0.34   &   0.41   &   0.25   &   0.54   \\
            10.0    &   0.40   &   0.46   &   0.28   &   0.58   \\
            14.7    &   0.46   &   0.51   &          &   0.60   \\
            21.5    &   0.55   &   0.61   &   0.45   &   0.72   \\
            31.6    &          &   0.78   &   0.67   &   0.76   \\
            \hline

 \end{tabular}
\end{center}

\end{table}

In the above, we have selected $\RL$ for test particles that allows
for reasonable comparison with fluid simulations. However, 
the dynamically important CRs in the ISM have much smaller $\RL$.
Since it appears that the time of the peak, $\tpeak$, and $\tau$ are connected,
an investigation of how $\tpeak$ varies with particle energy (or $\RL$)
may suggest a suitable range of $\tau$ to use in more realistic fluid
simulations. In this section, we perform test-particle simulations, but now 
utilizing the \emph{continuum model} for the
magnetic field described by \citet{Snodin2016},
where the magnetic field is constructed in the form of Fourier series
with random phases. This implementation allows one to probe
much lower particle energies than the discrete mesh model used above,
where the mesh spacing controls the smallest value of $\RL$ accessible.
\red{However, it still remains impractical to faithfully
explore energies much below $\RL \lesssim 10^{-4}L$, and so one can at best
extrapolate to infer the behaviour at the smallest relevant scales}.
This model magnetic field has fluctuations at a wide range of 
geometrically-spaced wave numbers to ensure sufficient resonant scattering of 
the particles at small scales.
We take $k_\text{max}^{\text{particle}}/k_\text{min}=1024$, and
$N=1024$ distinct wave vectors.

In Fig.~\ref{fig:particleTpeak} (data shown in Table~\ref{tab:particleTpeak}, we show the dependence of $\tpeak$ on $\RL$ and on the spectral index, $s$, using a set of test particle
simulations. The scaling with $\RL$ is approximately linear in all cases.
At sufficiently large $\RL$ ($\RL \gtrsim 0.03L$ for the values of $s$ 
considered here),
the peaks vanish (i.e. the peak and asymptotic values coincide), so $\tpeak$ cannot be determined.
Apparently,  there is a non-zero intercept at low $\RL$,
which appears to depend linearly on $s$.

At such small scales, it is expected that particles will be scattered by self-generated waves
\citep[e.g.,][]{KP69,Wentzel74}, which may have some effect on $\tpeak$.
In an attempt to account for this we adopt the simple scattering model used
by \citet{SetaEA2018}, which introduces an isotropic random change to the particle pitch
angle with respect to the local magnetic field once per Larmor time $t_{\rm L}$. In this model, the average change in the angle is proportional to $(R_L/L)^{1/4}$.
Applying this to the $s=-5/3$ case, we find a reduction in the intercept, but 
not a significant difference in the slope, as shown in 
Fig.~\ref{fig:particleTpeak}.

\section{Discussion and Conclusions}
\label{sec:conclusion}

Inspection of Fig.~\ref{fig:variance} shows that the Fickian prescription for the
diffusion of cosmic rays is unable to capture the transition from a ballistic to a
diffusive regime, which is manifest at earlier times in test particle simulations.
Nevertheless, the effective diffusivity asymptotically converges to that of the test
particles using $\kpara$ and $\kperp$ as calibrated for the Telegraph 
prescription.

The detailed agreement between the test-particle results and solutions of
the telegraph equation in Fig.~\ref{fig:slices} is remarkable, given that the
diffusion coefficients adopted are global constants that have no direct 
dependence on the local magnetic field properties. This suggests strongly that
the telegraph equation is a viable model for simulations of cosmic ray
propagation using a fluid description, with a subgrid model for their 
diffusion. The higher is the spatial resolution of such simulations and the 
larger is the particle energy, the more important can be initial deviations 
of the particle propagation from a diffusive behaviour at the numerically 
resolved scales, and the properly calibrated telegraph equation proves to be
able to capture such behaviours.

Our results lead to an  estimate of the time scale $\tau$, a parameter in the
telegraph equation.
Taking the outer scale of the turbulence $L\approx100\pc$ and assuming that
the dynamically important CRs (mainly protons) have energy of order $1\GeV$
(i.e., $R_L\simeq 10^{12}\cm$), the presence of a nonzero intercept in
Fig.~\ref{fig:particleTpeak} indicates that 
$\tau \simeq \red{20}\yr$ -- under the assumption of $\tau\approx0.5\tpeak$ and
taking ${\tpeak\approx \red{0.12}\times(100\pc)/c}$\red{,
extrapolated from the self-scattering ${s=-5/3}$ case.}
The maximum propagation speed for disturbances in the CR fluid is related to
$\tau$ and follows as ${v_\text{max}=(\kpara/\tau)^{1/2}\simeq 10^9\cm\s^{-1}}$.

It is worth noticing that the asymptotic value of the global diffusivity, $\keff$, is
about \emph{an order of magnitude smaller} than the local diffusivity, $\kpara$, which
is applied at the mesh scale in the fluid simulation.
This must be taken into account when choosing \red{simulation} parameters:
\red{
to achieve a global effective diffusivity of order $10^{28}\cm^2\s^{-1}$ ---
consistent with best fits to CR observational data \citep{SMP07} ---
our results suggest that a local diffusivity of at least
$\kpara\sim10^{29}\cm^2\s^{-1}$ needs to be adopted.
}

We have assumed CRs 
are scattered primarily by extrinsic
turbulence, rather than self-generated waves that would lead to 
self-confinement \citep[see, e.g.,][for a discussion of these two 
regimes]{Zweibel2017}. (For self-confined CRs, one might employ modified fluid 
equations, such as those of \citealt{Thomas2018}.)
We have also neglected dynamical effects of CRs on the gas density and velocity 
and, thereby, on the magnetic field. 
It is, however, unlikely that such effects can change the nature of
the CR diffusion process.

\section*{Acknowledgements}
\red{We thank the referee for their suggestions.}
We thank Amit Seta and Sergei Fedotov for useful discussions.
LFSR, GRS and AS acknowledge financial support of STFC (ST/N000900/1, Project 2)
and the Leverhulme Trust (RPG-2014-427).
APS was supported by the Thailand Research Fund (RTA5980003).
This research has made use of NASA's Astrophysics Data System.
{
\footnotesize
\noindent
\bibliographystyle{mnras}
\bibliography{fickian}
}

\appendix

\label{lastpage}
\end{document}